\documentstyle[pra,aps,epsf,twocolumn]{revtex}

\newcommand{\ket}[1]{\,|#1\rangle}
\newcommand{\sandwich}[3]{\langle #1|#2|#3\rangle}
\newcommand{\sumInt}[1]{{\sum \!\!\!\!\!\!\!\! \int_{#1} \  }}

\begin{document}
\draft 
\title{Interference effects in two-photon ATI by multiple
  orders high harmonics with random or locked phases}

\author{Eric Cormier$^{1,2}$, Anna Sanpera$^{1,3}$, Maciej
  Lewenstein$^{1,4}$ and Pierre Agostini$^{1}$} \address{$^{1}$Service
  des Photons, Atomes et Mol\'{e}cules, DRECAM, \\ Centre d'Etudes de
  Saclay,\mbox{F-91191 Gif-Sur-Yvette, France.} \\ $^{2}$CELIA, \\ 
  Universit\'{e} de Bordeaux I, 351 Cours de la Lib\'{e}ration,
  \mbox{F-33405 Talence, France.}\\ $^{3}$ Departament de F\'\i sica.
  Universitat Aut\'onoma de Barcelona, \mbox {08193 Bellaterra,
    Spain.}\\ $^{4}$ Institute f\"ur Theoretische Physik,
  Universit\"at Hannover, \mbox{30167 Germany.} } 

\date{\today}
\maketitle

\begin{abstract}
  We numerically study 2-photon processes using a set of harmonics
  from a Ti:Sapphire laser and in particular interference effects in
  the Above Threshold Ionization spectra. We compare the situation
  where the harmonic phases are assumed locked to the case where they
  have a random distribution. Suggestions for possible experiments,
  using realistic parameters are discussed.
\end{abstract} 


\narrowtext

\section{Introduction}
\label{sec:introduction}

High Order Harmonic generation (HOHG) refers to the radiation emitted
during the interaction of a short laser pulse with  a gas
jet. The spectrum of such a radiation consists of a series of odd
harmonics of the fundamental frequency having approximately equal
conversion efficiency (plateau), and followed by an abrupt cut-off. A
complete description of harmonic generation which allows for a direct
comparison with the experiments should include not only the response
of the single atom to the laser field, but also the propagation of the
different harmonics generated through the macroscopic
medium. Recently, several studies on harmonic generation have focused
on the properties of the phase of the generated harmonic waves
considering both, single atom and propagation
effects\cite{Salieres:1998:SST}. Within the frame of the single atom
response the strength and the phase of each harmonic wave corresponds to
the modulus and the phase acquired by the dipole moment $d(t)$, or
equivalently, the dipole acceleration induced by the external laser
field. Thus the spectrum of the radiated harmonics can be directly
calculated from the Fourier transform of the dipole acceleration\cite
{Schafer:1997:U}:
\begin{equation}
D(\omega)=\int \ddot{d}(t) e^{i\omega t} dt= A(\omega) e^{i \phi(\omega)},
\end{equation}
where $A(\omega)$ is the amplitude, $\phi(\omega)$ is the phase of the
dipole acceleration in the frequency domain, and $|D(\omega)|^2$ the
strength of the harmonics. The single atom response shows that the
phase of each harmonic, $\phi(\omega)$, is usually shifted with
respect to the fundamental phase and depends strongly on the intensity
of the laser field\cite {Lewenstein:1995:U}. Let us recall that the
harmonics are said phase-locked if the phase of each individual
harmonic is a linear function of its order {\em q} so it can be
written as $\phi_q = \beta q \omega + \zeta$ where $\beta q \omega$ is
the linear $q$-dependent term and $\zeta$ a constant term identical
for all harmonics. In general, harmonics from the plateau region
present a random phase when compared to each other, and only the
harmonics from the cut-off region are locked in phase. This can be
intuitively understood using the semi-classical description of the
atom-field interaction in which harmonic generation results from the
recombination of those electrons which have previously tunneled out
through the effective potential produced by the Coulomb and the laser
field. The induced dipole, and hence the process of harmonic
generation corresponds to trajectories in which the electron returns
to the core with appropriate kinetic energy and recombines emitting a
photon. Each contribution contains then a phase factor equal to the
real part of the action acquired by the electron following the
respective trajectory\cite{Lewenstein:1995:U}. Among all possible
trajectories, very few of them are relevant. For the harmonics in the
cutoff there is only one dominant trajectory, corresponding to a
single recombination time and, therefore, the phase of the harmonics
in this region remains locked. On the other hand, for the harmonics
lying in the plateau, there are typically two relevant trajectories
corresponding to two different returning times. The phase of such an
harmonic contains the contributions of these two trajectories and in
particular their quantum interferences. This quantum interefence term
leads to an apparently random phase. In spite of this fact, when one
analyzes the harmonic emission in time rather than in frequency
domain, one observes a clear periodicity of the signal. Antoine et al.
\cite{Antoine:1996:U} have shown that the time dependent emission
consist of a train of ultrashort pulses with two dominant pulses per
half cycle, each corresponding to one of the relevant trajectories.
Further studies show that under certain geometrical conditions it
should be possible to phase-match the contribution of only one of
these trajectories. That results in a train of ultrashort pulses
equivalent to those obtained by combining several plateau harmonics
with their phases locked.

 From an experimental point of view, determining the relative phase of
the emitted harmonics is not a straightforward task. Its study
necessarily involves some type of interference effects between the
different harmonics.  Although some pioneering work on interference
between harmonics of the same order have been recently
reported\cite{Zerne:1997:PLH,Ditmire:1998:U,Bellini:98}, a conclusive result on the
relative phase has not yet been achieved.

In this paper we address this question using a combination of
successive odd harmonics to study two-photon processes. When several
harmonics irradiate simultaneously an atom, there exist various paths
associated with 2-photon transitions ending up into the same final
state thus leading to possible interference effects. Those effects
should be clearly present in the ionization rates, as well as in the
photoelectron spectra. Similar effects of sensitivity to harmonics phases
have been previously studied in a series of papers on ATI by mixing 
harmonics and infrared laser fields\cite{Veniard:95,Veniard:96}.  
Furthermore, two-photon ionization of atoms
with photons pertaining to the VUV-XUV domain have been recently
reported for argon using the third harmonic of KrF laser
\cite{Xenakis:1996:OTI}, and for helium using the 9$^{{\rm th}}$
harmonic of a Ti:Sapphire laser\cite{Kobayashi:1998:EUP} giving thus
evidence of its experimental feasibility. In this paper, we focus on
two-photon ionization of atoms by Ti:Sapphire harmonics whose order
varies from $q=11$ to $q=33$.

Essentially, the aim of the paper is to show that measurable
quantities such as ionization or the more detailed photoelectron
spectrum have a different behavior depending on the relative phases of
the harmonics and to study these effects quantitatively. The paper is
organized as follows: in Section \ref{sec:Combi} we develop a
combinatorial model derived from perturbation theory which is able to
provide rough scaling laws for the total ionization rates as a
function of the number of harmonics present in the field.  Section
\ref{sec:ATI} reports on numerical  Above Threshold Ionization rates
obtained  by solving the Time-Dependent Schr\"{o}dinger Equation
(TDSE) with realistic polychromatic pulses for atomic H. Section
\ref{sec:MPI} is a study of quasi-resonant multiphoton ionization
(MPI) in He$^+$ also obtained by solving TDSE. Finally, in Section
\ref{sec:conclusion} we give concluding remarks and we comment on
possible experiments.

\section{A combinatorial model}
\label{sec:Combi} 
 
The model presented in this section provides an estimate for the
ionization probability as a function of the number of successive odd
harmonics included in the laser field. It is based on a 
combinatorial approach and a perturbative treatment of the atom-field
interaction.

\begin{figure*}[hf]
\epsfxsize=8cm \epsfbox{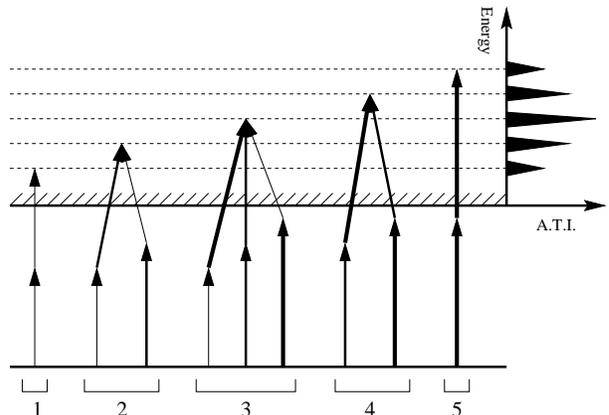}
\caption{Sketch of the processes investigated in the case of the
mixing of three harmonics. Also shown is the corresponding electron
spectrum.}
\label{fig:ATI_fig} 
\end{figure*}

Let us first sketch the problem under investigation. We consider
electrons emitted during ionization via a 2-photon process from an
atom initially in its ground state (the possible case where some
excited intermediate state come into resonance with the absorption of
only one photon is discussed in Section \ref{sec:MPI}). Since the
field considered here consists of  a linear combination of successive
odd harmonics:
\begin{equation}
E(t)=\sum_q E_{q}cos(q\omega _{L}t+\phi _{q})
\end{equation}
photons of different energies will be involved in the production of
electrons with different kinetic energies, as it is illustrated on
Fig. \ref {fig:ATI_fig}. Note that the energies at which electrons are
released appear on the photo-electron spectrum separated exactly by
the energy of 2 photons of the fundamental field $(\omega_L)$ which
generates the harmonics.  Moreover, electrons at a particular energy
may have been released by the absorption of photons originated from
distinct harmonics. The only restriction being that whatever the
combination of photons is, the sum of their energies ought to be the
same. For example, the production of electrons in the central peak of
the electron spectrum of Fig. \ref{fig:ATI_fig} results from
transitions following three different quantum paths. The amplitudes
associated to each quantum path may interfere depending on the
relative phase of the photons involved. The interference pattern
appears clearly if one writes the ionization probability associated
with these processes (labeled 3 on Fig.
\ref{fig:ATI_fig}) within perturbation theory:
\begin{eqnarray}
\label{eq:ionization_prob} 
p_{fg}=| 
&&^{a}M_{fg}^{(2)}E_{1}E_{3}e^{i(\phi _{1}+\phi _{3})}+ \nonumber \\
&&^{b}M_{fg}^{(2)}(E_{2}e^{i\phi _{2}})^{2}+ \nonumber \\
&&^{c}M_{fg}^{(2)}E_{3}E_{1}e^{i(\phi _{3}+\phi _{1})}|^{2}, 
\end{eqnarray}
where the subscripts $3$,$1$ and $2$ refer to the biggest, the
smallest and the intermediate photon, $E$ refers to the maximum field
amplitude, and $ \phi $ refers to the phase of the photons. The transition
amplitude matrix elements from the initial state $\ket{g}$ to the
final state $\ket{f}$ for the three paths are given by:
\begin{eqnarray}
\label{eq:transition_amplitudes}
^{a}M_{fg}^{(2)} &=&\lim_{\epsilon \rightarrow 0}
\sumInt{n}\frac{\sandwich{f}{z}{n}\sandwich{n}{z}{g}}{
\omega _{g}+\omega _{1}-\omega _{n} + i\epsilon},\nonumber   \\
^{b}M_{fg}^{(2)} &=&\lim_{\epsilon \rightarrow 0}\sumInt{n}\frac{\sandwich{f}{z}{n}\sandwich{n}{z}{g}}{
\omega _{g}+\omega _{2}-\omega _{n} + i\epsilon},  \nonumber \\
^{c}M_{fg}^{(2)} &=&\lim_{\epsilon \rightarrow 0}\sumInt{n}\frac{\sandwich{f}{z}{n}\sandwich{n}{z}{g}}{
\omega _{g}+\omega _{3}-\omega _{n} + i\epsilon},  
\end{eqnarray}
where the summation runs over all coupled intermediate states
including the continuum. The special case of ATI can be numerically
extrapolated\cite{Cormier:1995:EM} or computed accounting explicitely
for the resonant free-free transition matrix
elements\cite{Cormier:1993:DT}. Since we assume the interaction with a
linearly polarized field and express it in the length gauge, only the
matrix elements of $z$ appear in expression
(\ref{eq:transition_amplitudes}). The ionization probability then
reads:
\begin{eqnarray}
\label{eq:proba_interferences} 
p_{fg} &=&
|^{a}M_{fg}^{(2)}|^{2}E_{1}^{2}E_{3}^{2}+
|^{b}M_{fg}^{(2)}|^{2}E_{2}^{4}+
|^{c}M_{fg}^{(2)}|^{2}E_{3}^{2}E_{1}^{2}+ \nonumber \\
&&
(^{a}M_{fg}^{(2)}\ ^{c}M_{fg}^{(2)}\ ^{*}+c.c.)E_{1}^{2}E_{3}^{2}+ \nonumber \\
&&
(^{a}M_{fg}^{(2)}\ ^{b}M_{fg}^{(2)}\ ^{*}E_{1}E_{3}E_{2}^{2\ *}e^{i(\phi_{1}+\phi
_{3}-2\phi _{2})}+c.c.)+ \nonumber \\
&&(
^{b}M_{fg}^{(2)}\ ^{c}M_{fg}^{(2)}\ ^{*}E_{2}^{2}E_{3}^{*}E_{1}^{*}e^{-i(\phi_{1}+\phi
_{3}-2\phi _{2})}+c.c.). 
\end{eqnarray}
Interferences in the probability given  by Eq. (\ref
{eq:proba_interferences}) are governed by the phase difference $\Delta
=\phi_{1}+\phi _{3}-2\phi _{2}$ that is, the relative phase between
the harmonics involved in this process.  Such a partial probability
can be derived for all accessible final states. Note, that
interferences are always absent in processes involving less than 3
different photons which is always the case for the processes leading
to the two first and two last final states (like processes labeled 
1,2,4 and 5 in Fig. \ref{fig:ATI_fig}) independently of the total
number of harmonics included in the field.

When the incident field is a linear combination
of successive harmonics, there exist several distinct quantum paths
associated to transitions to the same final state. These paths may
interfere depending on the relative phase between the photons. Obviously,
the number of possible interfering paths increases with the number of
different frequencies included in the field.

To simplify the problem, we will now assume that all harmonics have
the same strength, that is $E_{1}=E_{2}=E_{3}=E$. As the
harmonics we are considering here all come from the plateau part of
the spectrum, this assumption is generally fulfilled. Furthermore, we
also assume that all 2-photon transition amplitude matrix elements (as
they are defined in Eq. (\ref{eq:transition_amplitudes})) are real and 
have the same value, i.e. $^{a}M_{fg}=^{b}M_{fg}=^{c}M_{fg}=M$ . Although
never fulfilled in general, this approximation is
reasonable if the first photon absorbed brings the system into a region free
of intermediate resonances. For the case of our illustrative example, the
ionization probability associated to the central process
 (labeled 3 in Fig. \ref{fig:ATI_fig}), corresponds to:
\begin{equation}
\label{eq:proba_interf_simplified} 
p_{fg} = M^{2}E^{4} (5 + 4\cos (\phi_{1}+\phi _{3}-2\phi _{2})).
\end{equation}
Let us now analyze  the two configurations of harmonic phases we are
interested in, that is, the phase locked case and the random
distribution of phases. As defined in the introduction, if harmonics are
phase locked to each other, their phase has the form: 
$\phi_q = \beta q \omega + \zeta$. Applied to
our  example 
$\Delta=\phi_{1}+\phi _{3}-2\phi _{2}$ simply reduces to
zero. Therefore, the cosine function in Eq.(\ref{eq:proba_interf_simplified}) takes value one. 
On another hand, we can interpret the  random distribution of phases as
a loss of  coherence between different quantum paths. It is then
reasonable to associate the random distribution to the  average value of  
Eq.(\ref{eq:proba_interf_simplified}) over all possible phases. 
Since the average value of the cosine function is zero, all
terms in the probability vanish except the constant
(phase-independent) terms. 
Generalizing these results for an arbitrary number $N$ of incident
harmonics, we obtain that the ionization probability associated to
the central peak behaves as:

\begin{eqnarray}  \label{eq:PL_central}
P^{(PL)}_c(N) & = & \alpha^2 N^2
\end{eqnarray}
and 
\begin{eqnarray}  \label{eq:RP_central}
P^{(RP)}_c(N) & = & \alpha^2 \left\{ 
\begin{array}{cl}
2N-1 & \mbox{if $N$ odd} \\ 
2N & \mbox{if $N$ even}
\end{array}
\right.
\end{eqnarray}
where PL refers to phase locked configuration, RP to random phase 
configuration and
$\alpha=ME^2$. The total 2-photon ionization probability is straightforwardly
obtained by
integrating the electron spectrum over the electron energy range
relative to the 2-photon processes. This probability accounts for the
side peaks in the photoelectron spectrum and is obtained by simply 
adding the probability
associated to each side peak (see Fig. \ref{fig:ATI_fig}). 
It can be shown that the total
probabilities read:
\begin{eqnarray}
P^{(PL)}(N) &=&\alpha ^{2}\left( 2\sum_{n=1}^{N-1}n^{2}+N^{2}\right) 
\label{eq:PL} \\
&=&\alpha ^{2}\left( \frac{2N^{3}}{3}+\frac{N}{3}\right)   \nonumber
\end{eqnarray}
and 
\begin{eqnarray}
P^{(RP)}(N) &=&\alpha ^{2}\left( 4\frac{N(N-1)}{2}+N\right)   \label{eq:RP}
\\
&=&\alpha ^{2}\left( 2N^{2}-N\right)   \nonumber
\end{eqnarray}
As mentioned earlier, when ordered by increasing final state energy,
the two first and two last processes will never show path
interferences independently of the total number of harmonics
involved. Also, the process having the largest number of interference
terms will always be the central one (like process number 3 in
Fig. \ref{fig:ATI_fig}). 

To summarize, the total ionization associated with 2-photon absorption
should vary approximately as $2/3 N^{3}$ if the harmonics are locked in
phase, and like $2 N^{2}$ if the relative phases between them is
random. Accordingly, the ionization rate associated to the central
process should vary like $N^{2}$ in the phase locked case and like $2N$
if the phase distribution is random.

We expect that similar scaling properties will characterize other
processes of interest, such as 2-photon Above Threshold Ionization
(ATI) which we shall discuss in the next section. The advantage of
studying ATI instead of 2-photon ionization processes is that such
processes are free from intermediate resonances that modify the
interference pattern, as we shall discuss in Section
\ref{sec:MPI}. The disadvantage on the other hand, is that the ATI
signal might be low and, consequently the effects due to the relative
phase of the harmonics might be difficult to observe experimentally.

Finally, it is worth stressing here that the problem considered in
this section has a lot in common with the problems of ionization by
stochastic laser fields, discussed in the literature in the 70's and
80's\cite{Shore}. In particular, the enhancement by a factor $K!$ of
the $K$-photon ionization probability by multimode laser fields has
been predicted and observed.  In our random phase model, the same
effect is responsible for the fact that $P^{(RP)}\simeq 2!N^2$. On the
other hand, the problem considered in this paper is more general and
deals with the differences induced by random and phase locked
configurations in laser pulses in {\it real} atoms, i.e. it fully
accounts for atomic bound states and continuum structures.

\section{2-Photon Above Threshold Ionization}
\label{sec:ATI}

In this section, ATI is investigated using a time-dependent 
non-perturbative approach. It consists in solving the Time-Dependent
Schr\"{o}dinger equation (TDSE) for a hydrogen atom interacting with
the combination of several harmonic fields. The method used to solve
the TDSE and to compute the electron spectrum is based on the
expansion of the total wave-function on spherical harmonics and
B-spline functions (see \cite{Cormier:1996:OGG}).  The temporal
propagation in the length gauge is well adapted to the rather low
intensities considered here. The electron spectra correspond to
electrons emitted in the direction of the polarization.

\subsection{Ionization of hydrogen in its ground state}

\label{subsec:H(1s)}

\begin{figure*}[hf]
\epsfxsize=8cm \epsfbox{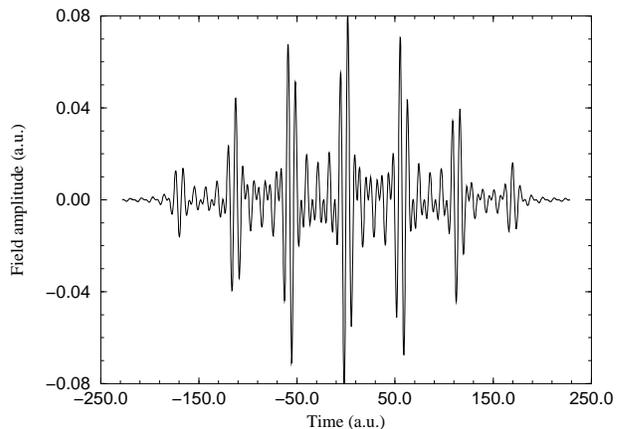}
\caption{Electromagnetic field amplitude corresponding to a linear
combination of 5 odd harmonics ranging from order 11 to 19
($\hbar\omega_L = 1.5$ eV).}
\label{fig:N=5_field} 
\end{figure*}

\begin{figure*}[hf]
\epsfxsize=8cm \epsfbox{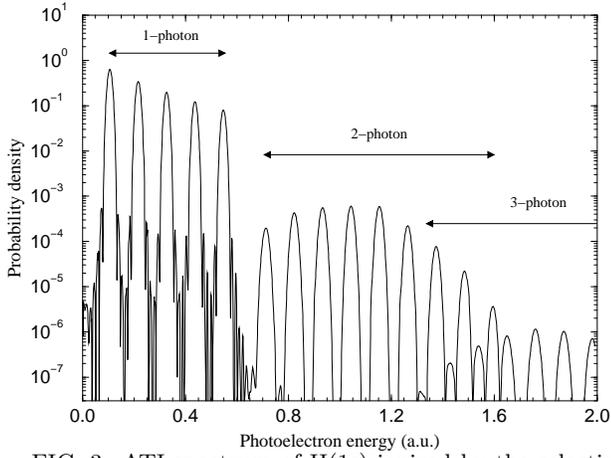}
\caption{ATI spectrum of H(1s) ionized by the selection of harmonics
defined in Fig. \ref{fig:N=5_field}. The 3 different subsets
of ATI peaks corresponds to different order of processes.}
\label{fig:ATI_N=5} 
\end{figure*}

We have computed the electron spectra obtained by shining a linear
combination of successive harmonics on atomic hydrogen in its ground
state.  The frequencies of the selected harmonics are chosen so that
absorption of a single photon already brings the system into the
continuum. Thus, the 2-photon processes we are interested in belong to
the so-called Above Threshold Ionization regime. In our approach, the
field is described semi-classically as:
\begin{equation}
E(t)=E_{o}f(t)\sum_{n=1}^{N}\cos (q\omega _{L}t+\phi _{q}),\qquad
q=2(n-1)+q_{o},  \label{eq:classical_field}
\end{equation}
where $f(t)$ is the normalized envelope of the field, in our case a
cosine squared with $5\ {\rm fs}$ full width at half maximum (20
cycles of harmonic 11); $E_{o}$ is the maximum field amplitude
(identical for all harmonics) corresponding to an intensity of
$10^{13}\ W/cm^{2}$; $\hbar \omega_{L}=1.5$ eV is the fundamental
photon energy and $q_o$ indicates the lowest harmonic component in the
polychromatic field. The number of harmonics we consider varies from
$N=2$ to $N=5$, so that the odd harmonics included are H13-- H15 for
$N=2$, H13--H17 for $N=3$, H11--H17 for $N=4$ and H11--H19 for $N=5$.
The field corresponding to the case $N=5$ is plotted in Fig. \ref
{fig:N=5_field}. The electron spectrum due to such a field is shown on
Fig. \ref {fig:ATI_N=5} where we observe that electrons produced via
different order processes are clearly distinguishable. The spectrum
has to be read as follows: there are 3 clear subsets of ATI peaks,
each of them corresponding to a different order process. The left most
subset corresponds to order 1, that is, direct photo-ionization via
the absorption of a single photon (H11 for the very first peak and H19
for the fifth). The second subset, which is the one investigated here
embodies all processes of order 2. For example, the third peak of that
subset results from the simultaneous absorption of either two photons
from the harmonic 13, or one from the harmonic 11 and another from the
harmonic 15. All possible allowed polychromatic combinations are
included in this type of simulation. Higher order processes appear
with exponentially decreasing probability and, therefore, their study
have a restricted interest from an experimental point of view.

We have limited ourselves to a maximum of 5 combined harmonics in this
study because higher values would lead to overlapping subsets of ATI
peaks as it is almost the case for $N=5$ where the 3-photon processes
subset partly overlap the 2-photon subset (see
Fig. \ref{fig:ATI_N=5}).

Let us now focus on the 2-photon Above-Threshold Ionization, that is,
the central subset of ATI peaks in Fig. \ref{fig:ATI_N=5}. The problem
of interest here is to determine how much does the total ionization
associated with 2-photon absorption increase as we increase the
number of harmonics in the field assuming that either all harmonics are
locked in phase, or  their relative phases are random.

\begin{figure*}[hf]
\epsfxsize=8cm \epsfbox{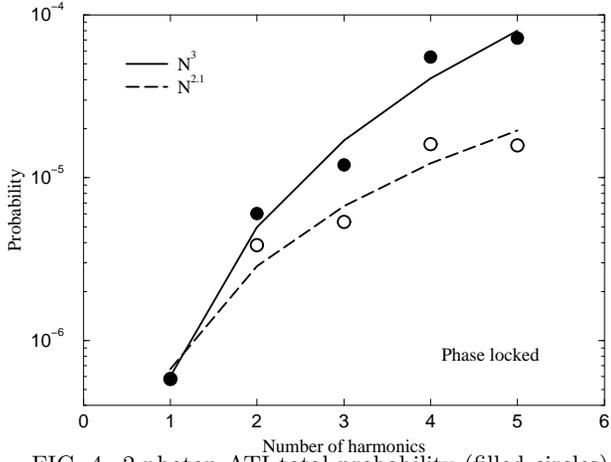}
\caption{2-photon ATI total probability (filled circles) and 2-photon
ATI probability restricted to the central peak (open circles) as a
function of the number of harmonics $N$ for the locked phase case. The line and dashed line are power law fitting
curves with a respective argument of 3 and 2.1.}
\label{fig:Power_Law_PL} 
\end{figure*}

We have first simulated the case of phase-locked harmonics. This is
achieved by solving the TDSE when all harmonic phases $\phi_q$ are set
to zero.
Fig. \ref {fig:Power_Law_PL} shows the 2-photon
ionization probability as a function of the number of harmonics
included in the field. The main result to be read from
Fig. \ref{fig:Power_Law_PL} is that ionization certainly scales like
the cubic power of the number of harmonics as predicted by the
combinatorial approach Eq.(\ref{eq:PL}). Similarly, the ionization
probability associated with the central ATI peak scales like
$N^{2.1}$, in good agreement with Eq.(\ref{eq:PL_central}).

\begin{figure*}[hf]
\epsfxsize=8cm \epsfbox{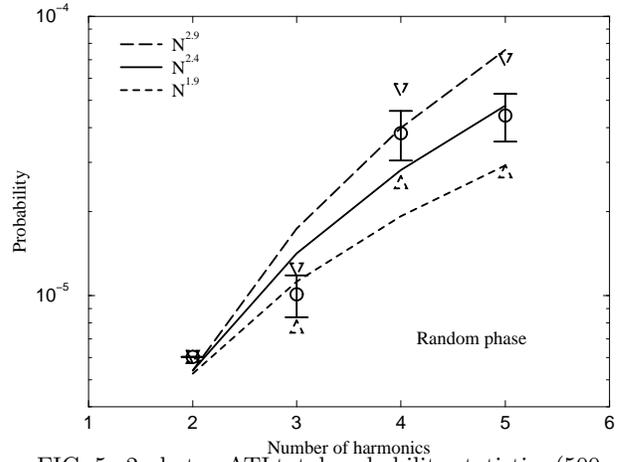}
\caption{2-photon ATI total probability statistics (500 runs) as a
  function of $N$. The relative phases are random. Open circles refer
  to the average value of the distribution (power law fit with an
  argument of 2.4. Also shown are the maximum, minimum values of the
  distribution and the standard deviation (Triangles down, up and
  error bars respectively).}
\label{fig:Power_Law_RP_TOT} 
\end{figure*}

\begin{figure*}[hf]
\epsfxsize=8cm \epsfbox{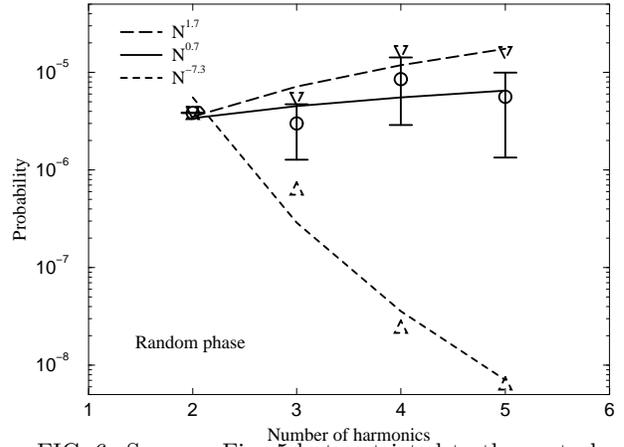}
\caption{Same as Fig. \ref{fig:Power_Law_RP_TOT} but restricted to 
the central peak. Open circles: average value. The power law fit gives an argument of 0.7.}
\label{fig:Power_Law_RP_Central} 
\end{figure*}

To simulate the case where the phases are random, we have accumulated 
the results of 500 different random phase configurations after solving the
Schr\"{o}dinger equation for each of them and we have averaged 
the results at the end. This has been done for each value of $N$.  
We report in Fig. \ref{fig:Power_Law_RP_TOT} the ionization probability thus
obtained together with its statistics fluctuations. The most
interesting result here is the behavior of the average value of the
ionization (open circles in Fig. \ref{fig:Power_Law_RP_TOT}). The
average value scales like $N^{2.4}$, departing significantly from the
combinatorial estimate that predicts a $N^2$ law
(Eq. \ref{eq:RP}). Let us now examine the statistical fluctuations. On
one hand, we find that the maximum value of the ionization (down
triangles in Fig. \ref{fig:Power_Law_RP_TOT}) scales as $N^{2.9}$,
very close to that of the phase locked case.  Obviously, the
configurations that lead to such maximal value are the special cases
where all phase differences are close to zero like in our 
simulation of the phase locked configuration.  
On the other hand, the most globally
destructive phase configuration leading to the minimum value of the
ionization probability scales like $N^{1.9}$ (up triangles in Fig.
\ref{fig:Power_Law_RP_TOT}). We have also investigated the ionization
probabilities associated to the central 2-photon ATI peak, since it is
the transition that involves the largest number of interfering quantum
paths. The results are presented in
Fig. \ref{fig:Power_Law_RP_Central}.  The average value of the
ionization probability scales as ($N^{0.7}$) in comparison with the
fit ($N^{2.1}$) obtained for the phase locked case.  Finally, its
minimum value drops dramatically since interferences now can be
completely destructive, that is, there exists a particular
configuration of phases able to totally suppress ionization for this
particular electron energy.  This feature is absent in the case of
total 2-photon ionization since, as explained before, the 4 outermost
ATI peaks correspond to transitions involving only two quantum paths,
and therefore are not affected by any change in phase.

A way to improve the statistical results is to increase the number of
harmonics included in the field by using higher order harmonics. This
will prevent us  from having different ATI order processes overlapping
but also the ionization signal will decrease because the cross-section
associated to these processes decays with increasing photon energy.

\subsection{Ionization of hydrogen in the 2s state}

\label{subsec:H(2s)}

A solution to ease 2-photon ionization due to higher harmonics is to
start from a less bound atomic energy level like H(2s) for example. 

By choosing harmonic 21 ($\hbar\omega_L = 1.5$ eV) as the
central component, we can combine up to 7 harmonics. Because the lower
binding energy, we set now the field intensity of all harmonic
components to $10^{10}$ W/cm$^2$. As expected with this intensity, the
probability for ionization by a 2-photon process is very low.
However, one should keep in mind that this probability varies like
$I^2$ in agreement with the process order, and that the intensity we
are using here underestimates experimental attainable fluxes.

\begin{figure*}[hf]
\epsfxsize=8cm \epsfbox{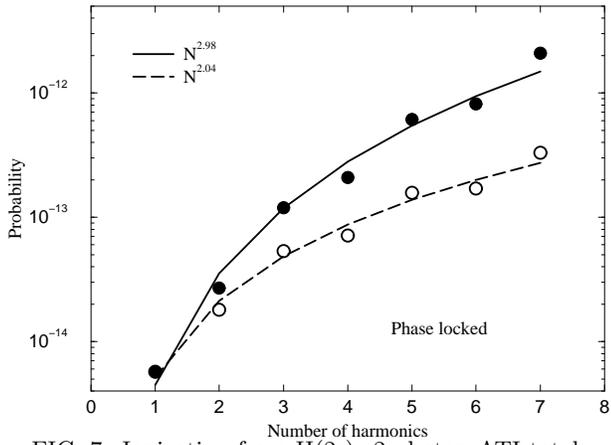}
\caption{Ionization from H(2s): 2-photon ATI total probability (filled
circles) and 2-photon ATI probability restricted to the central peak
(open circles) as a function of $N$. The relative phases are locked
(zero). The line and dashed line are power law fitting curves with a
respective value of 2.98 and 2.04.}
\label{fig:H2s_Power_Law_PL} 
\end{figure*}

We find that in the phase locked case, the scaling laws concerning
total ionization associated with 2-photon absorption shown in
Fig. \ref {fig:H2s_Power_Law_PL} agree very well with the
combinatorial estimates expressed in Eqs. (\ref {eq:PL_central}) and
(\ref{eq:PL}).

\begin{figure*}[hf]
\epsfxsize=8cm \epsfbox{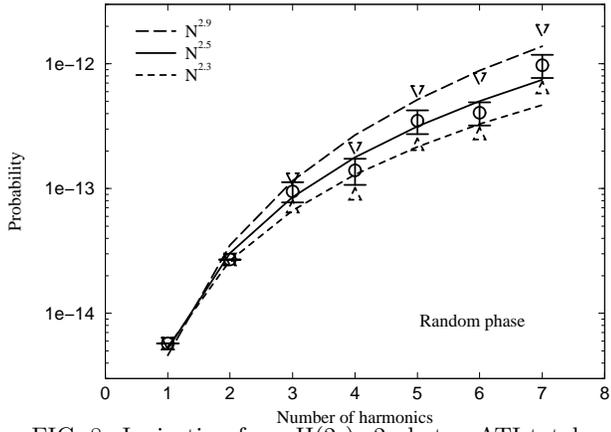}
\caption{Ionization from H(2s): 2-photon ATI total probability
statistics (300 runs) as a function of $N$. The relative phases are
random. Open circles refer to the average value. The power law fit
gives an argument of 2.5. Also shown are the maximum, minimum values
and the standard deviation (Triangles down, up and error bars).}
\label{fig:H2s_Power_Law_RP_TOT} 
\end{figure*}

The random phase simulation for H(2s) has been carried out by
accumulating results from 300 resolutions of the TDSE, each time with
a new random distribution of harmonic phases. In Fig. \ref
{fig:H2s_Power_Law_RP_TOT}, we show the 2-photon ATI probability,
obtained by integrating the spectrum over the whole
range of electron energies corresponding to 2-photon ATI processes. In
the case $N=7$, the integration runs from $E_k=40$ eV up to $E_k=76$
eV. Here again, there is a significant departure of the
average value of the ionization probability, which scales as
$N^{2.55}$, from the $N^2$ combinatorial law associated with the
random phase configuration.

Finally, as in the previous section, we have looked at the most phase
sensitive ATI peak located at the center of the 2-photon ATI subset in
the electron spectrum. The expected combinatorial power laws are 2 and
1 for harmonic phases, locked and random respectively. The quantum
mechanical computation gives the values of 2 and 1.3 for the respective
configurations  as it is  shown in
Fig. \ref{fig:H2s_Power_Law_RP_Central}. The lower limit of the
distribution drops abruptly until it reaches the background induced by
the numerical accuracy (around $10^{-16}$ on Fig. \ref
{fig:H2s_Power_Law_RP_Central}).

To summarize, we have shown that there is a different behavior in the
2-photon ionization probabilities as a function of $N$ if one
considers the harmonics locked in phase or if the distribution of
their phases is random. To a rough extend, scaling laws can be
explained using the idealized combinatorial model of Section II.

\section{2-photon ionization}
\label{sec:MPI}

The previous section has been devoted to the study of 2-photon ATI
processes, where absorption from a single photon already brings the
system into the continuum. Now, we propose to study direct 2-photon
harmonic ionization (MPI). In contrast with the 2-photon ATI cases,
ionization requires now two harmonic photons, so the interference effects
governed by the harmonic phases appear directly in the first subset of
peaks in the electron spectra.

At this point, let us recall that our combinatorial model is based on
the assumption that the total transition amplitude towards a given
final state is only related to the number of interfering paths leading
to such final state. This assumption, in turn, implies that all the matrix
elements involved in the transitions are considered equal. The results
obtained so far indicate that this is a reasonable assumption for
2-photon ATI processes where bound-bound transitions do not play any
essential role. However, 2-photon ionization processes (MPI) may
involve quasi-resonant bound-bound transitions. At low intensities,
i.e. in the frame of perturbation theory, the values of the
bound-bound transition matrix elements depend strongly on how close
the transition is from a resonance. As we shall see, this fact
substantially modifies the results we have encountered so far.

In order to optimize the study of the interferences between different
harmonics having, at the same time, a good control of the role of the
resonances involved, we choose as a model a hydrogenic ion (He$^{+}$).
Note, however, that the results we present in this section can be
straightforwardly generalized to any atom or ion and, therefore, have
a certain generic validity.
 
We numerically solve the TDSE for direct 2-photon ionization, where
the external field is a linear combination of different harmonic
fields, all of them with the same polarization and intensity.  As in
the previous sections, we consider harmonics from a Ti:Sapphire laser
($\hbar\omega_L=1.5$ eV), set the intensity of each harmonic to
$1.2\times 10^{12} $ W/cm$^2$ and calculate ionization probabilities
as well as photoelectron spectra for phase locked and random phase
configurations. For our choice of parameters, the smallest harmonic
from Ti:Sapphire required for 2-photon detachment of He$^{+}$ is the
19th and the largest one the 35th. Larger harmonics will directly
detach the ion, and smaller ones than 19th cannot achieve 2-photon
detachment.

\begin{figure*}[hf]
\epsfxsize=8cm \epsfbox{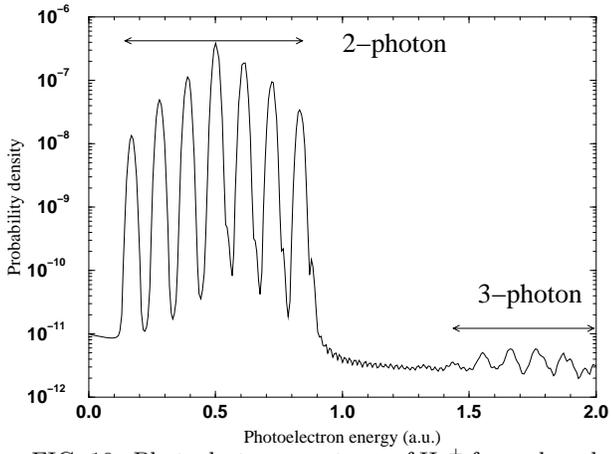}
\caption{Photoelectron spectrum of He$^+$ for a phase locked
configuration containing 4 harmonics (from 19th to 25th). The first
subset corresponds to direct 2-photon ionization processes. At higher
energies a second subset of peaks corresponding to 3-photon process
appears.}
\label{fig:D2P1} 
\end{figure*}

\begin{figure*}[hf]
\epsfxsize=8cm \epsfbox{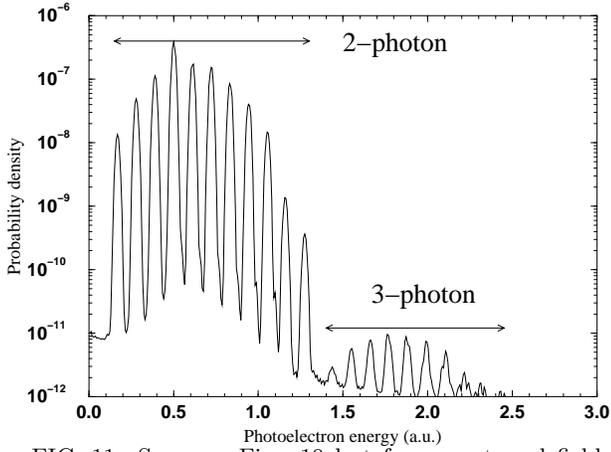}
\caption{Same as Fig. \ref{fig:D2P1} but for an external field
containing now 6 harmonics components (from 19th to 29th). Observe
that the peak distribution in the 2-photon ionization processes does
not follows the same pattern as before.}
\label{fig:D2P2} 
\end{figure*}

Before proceeding further, it is illustrative to show some of the
effects due to atomic resonances in the photoelectron spectrum.  To
this aim we display the spectrum for a phase locked configuration
containing: (a) N=4 harmonics ranging from 19th to 25th
(Fig.\ref{fig:D2P1}) and; (b) N=6 harmonics ranging from 19th to 29th
(Fig.\ref{fig:D2P2}). In both spectra there are two distinguishable
set of peaks: the first one corresponds to 2-photon detachment
processes, the second one corresponds to 3-photon processes (ATI
peaks).  From an experimental point of view, the interest in the
second subset should be restricted due to the low probability of
3-photon processes.
 
The first subset contains $2N-1$ different peaks, where $N$ is the
number of harmonics present in the external field. As mentioned in the
previous sections, the central peak of this subset corresponds, in
both cases, to the transition involving the largest number of
different quantum paths. While in Fig. \ref{fig:D2P1} this peak is the
most probable one, in Fig. \ref{fig:D2P2} the central peak is clearly
suppressed in comparison with lower order peaks.  The reason for this
peak suppression is due to the fact that the 27th harmonics (only
present in case (b)) crosses the $1s-2p$ resonance quasi-resonantly.

\begin{figure*}[hf]
\epsfxsize=8cm \epsfbox{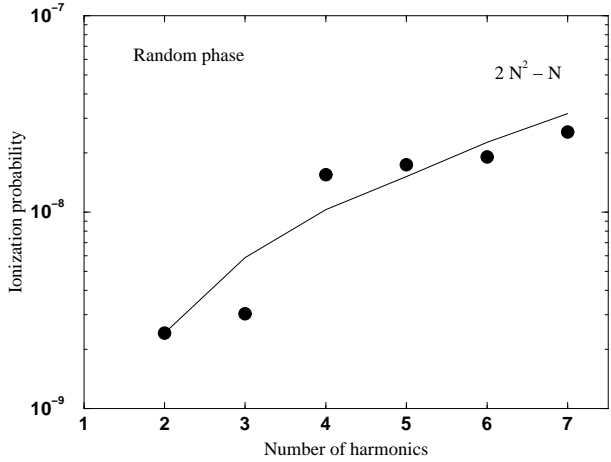}
\caption{Average value of the ionization probability versus the number
of harmonic components for random phase configuration. 
Full line corresponds to the combinatorial estimate (Eq.(\ref{eq:RP}))
and dots display the numerical results.}
\label{fig:D2P3} 
\end{figure*}

Let us now focus on the results corresponding to ionization 
probabilities. As in section \ref{sec:ATI}, the 2-photon detachment
probability is calculated by integrating the area below all 2-photon
processes in the photoelectron spectrum. We start by analyzing the
random phase configuration where  
we have accumulated results from a large number of random phase distributions
and we have averaged them at the end. In Fig. \ref {fig:D2P3} we plot the
average value of the ionization probability versus the number of incident
harmonics $N$ which varies from $N=2$
to $N=7$. The lowest harmonic involved is always the 19$^{{\rm th}}$ so $N=2$ 
includes 
harmonics 19-21, $N=3$ includes harmonics 19-21-23, and so on. The overall
behavior of the ionization probability obtained numerically (dots)
agrees relatively well with the combinatorial approach (full line).

\begin{figure*}[hf]
\epsfxsize=8cm \epsfbox{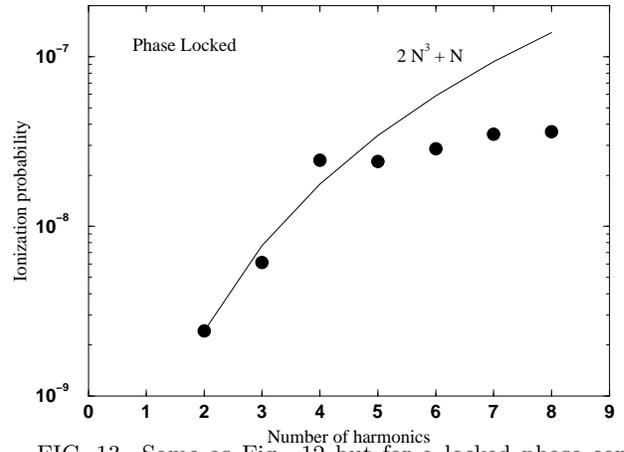}
\caption{Same as Fig. \ref{fig:D2P3} but for a locked phase configuration.
Full line corresponds to the combinatorial estimate (Eq.(\ref{eq:PL}))
and dots display the numerical results.}
\label{fig:D2P4} 
\end{figure*}

In Fig. \ref{fig:D2P4} we
present the corresponding ionization probability as a function of the
number of harmonics $N$, for the phase locked configuration. In the
simulation, 
all the individual harmonic phases, $\phi _{q}$,
have been set to zero. 
Again, the full line represents the results from
the combinatorial approach (Eq. \ref{eq:PL}), while symbols show our
numerical results. We observe that for low order incident harmonics
i.e. from 19th to 25th, the numerical results
agree relatively well with the combinatorial approach. However, when
higher harmonics are included (from 27$^{{\rm th}}$ onwards) the
combinatorial approach clearly overestimates the ionization
probability obtained numerically. The discrepancy between the estimate
and the numerical value becomes then dramatic.

This discrepancy can be understood by analyzing the role of resonances
within perturbation theory. Under the parameters we have used in the
simulations, the energy of the 27$^{{\rm th}}$ harmonic is slightly
detuned (from above) from the $1s-2p$ resonance. As  a result, the
matrix elements corresponding to transitions of the type:
\begin{equation}
M\propto\sumInt{n}\frac{\sandwich{f}{z}{n}\sandwich{n}{z}{1s}}{\omega
_{1s}+\omega _{27}-\omega _{n}} E_{27}*E_{i} \label{Eq:sign}
\end{equation}
where $i$ refers to the second photon absorbed, flip their sign
compared to those transitions which are far off resonance.

Let us analyze in more detail the 2-photon MPI transitions. From the 2
possible paths starting from the ground state 
leading to photodetachment; $s\rightarrow p\rightarrow s$
and $s\rightarrow p\rightarrow d$, the second one clearly dominates
over the first one. This is because transitions that increase their
principal quantum number and their angular momentum in the same
direction are always favored compared to the others
\cite{Bethe:1957:QMA}. Therefore, we focus on 2-photon transitions of
the type $s\rightarrow p\rightarrow d$ and present the values of 
$M$  in table \ref{tab:elements}

The table has to be interpreted as follows: rows refer to the first
absorbed photon and columns to the second one. For example, the first
row displays the values of the matrix elements corresponding to
transitions in which the first absorbed photon is the 19$^{{\rm th}}$
harmonic and the second one the one indicated by the column label,
i.e.  transition amplitude of the type:
$\sum\frac{\sandwich{f}{z}{n}\sandwich{n}{z}{1s}}
        {\omega_{1s}+\omega_{19}-\omega_{n}} E_{19}E_i$.

As it should be, we observe that the values of the matrix elements
 (i) decrease as the
energy of the final state increases and (ii) decrease for transitions
far from resonances. Immediately after a resonance has been crossed,
the matrix element flip its sign.  Thus,  with the parameters we
have use in the simulation, all matrix elements
corresponding to the absorption from the ground state of a photon from
the 27$^{{\rm th}}$harmonic (Eq. (\ref{Eq:sign})) flip their sign
compared to all the other matrix elements shown in the table.

This relative sign flip strongly affects the total ionization
probability in the phase locked case, for which $P^{(PL)}\propto |\sum
M_i |^2$, i.e.  proportional to the square of the sum of the
individual matrix elements. It has also consequences for random phase
configurations but there the effects are less dramatic.
 It is easier
to understand the difference between both cases with an
example. Consider an incident field containing harmonics 
25$^{{\rm th}}$, 27$^{{\rm th}}$and
29$^{{\rm th}}$. Such an incident field, leads to five possible 
final electron energy states,
which ordered by increasing energy correspond to: (i) 25+25
(absorption of 2 photons from harmonic 25$^{ {\rm th}}$); (ii)
$25+27$, $27+25$; (iii) $25+29$, $27+27$, $29+25$; (iv) $27+29$,
$29+27$ and finally (v) $29+29$. Remember that in order to have
interference effects we need at least 3 different quantum paths
contributing to the process. Therefore, the partial ionization
probabilities associated to states (i), (ii), (iv) and (v) are the
same regardless the relative phase of the harmonics. The partial
ionization probability associated to state (iii) which contains
contributions from 3 different quantum paths depends on the relative
phases of the harmonics. For phase locked harmonics, it is proportional
to 
$|M_{(25+29)}+M_{(27+27)}+M_{(29+25)}|^2$. On the other hand, if the
harmonics have a random phase, the partial ionization probability
associated to the final state (iii) is proportional to
$|M_{(25+29)}+M_{(29+25)}|^2+ |M_{27+27}|^2$ . Comparing both
expressions, we note that it is only in the phase locked case where
the sign of $M_{(27+27)}$  plays a definite role since it acts
effectively as a destructive interference term. Notice that the
values of the matrix elements corresponding to transitions
$27+25$ and $ 27+29$ do  modify the partial ionization 
probabilities associated to
the final states (ii) and (iv) respectively, but they do it 
exactly in the same
way for both cases: locked and random phase configurations.

To corroborate our findings we have extended our numerical results up
to the $33^{{\rm rd}}$ harmonic ($N=8$). For the parameters we have
used, the photon energy of harmonic 33$^{\rm rd}$ is slightly higher
than the atomic transition $1s\rightarrow 3p$. Consequently, the
matrix elements corresponding to absorption from the ground state of a
photon from harmonic 33$^{\rm rd}$ will again flip their sign,
 and again act
as an effective destructive interference for locked
phase configurations. As a result,  the disagreement between 
the ionization probability and the combinatorial estimate
should increase. Our numerical results support this conclusion.
 
In spite of the discrepancy between our numerical results and the
combinatorial estimates, it should be stressed here that nevertheless
the overall behavior of the ionization probability for random and
phase locked configurations remains distinct, allowing in principle
for a distinction between both cases.

We would like to remark here that accordingly to what we have seen, it
might also occur that for a {\em particular} final state energy, its
associated partial transition probability becomes larger for incident
harmonics with random phases than for phase locked harmonics.  Such a
partial peak inversion demonstrates that the effects due to atomic
resonances can be in some cases strong enough to compensate the
effects due to quantum phase interference.

To summarize, we have shown that that the overall behavior of the
ionization probability clearly depends on the relative phase
configuration of the incident harmonics. Furthermore, we have shown
that for phase locked configurations and incident harmonic fields
near-resonance, the total ionization probability strongly departs from
the combinatorial estimate and it does no longer follow the $N^3$
power law (Eq. \ref{eq:PL}).  If, on the contrary, the incident
harmonics are all far of atomic resonances, the combinatorial estimate
predicts accurately the exact results (Eq. \ref{eq:PL} and
\ref{eq:RP}).  We stress, however, that most two-photon ionization
experiments involving a set of harmonics is bound to show the effects
of resonances.  Finally, for incident harmonics with random phases,
the average value of the ionization probability agrees relatively well
with the combinatorial approach i.e.$N^2$.
\label{sec:conclusion}

\section{Conclusions}
In conclusion, we have shown that in 2-photon ATI of hydrogen atoms by
multiple orders high harmonics, measurable quantities such as the
ionization yield or the photoelectron spectra significantly depend on
the relative phase of the harmonics. It should therefore be possible
to determine experimentally the phase configuration of the high
harmonics and, in particular if the phases are locked. The method
suggested by the calculation is a measurement of these quantities as a
function of the number of selected harmonics. Clearly this is a
difficult task.  One difficulty lies in the relatively low photon
harmonic flux currently achievable: $10^{9}$ photons using $100\ fs$
lasers in argon, and $10^{7}$ photons using $30\ fs$ in neon
\cite{Salieres:1998:SST}.  Such a flux is in general too low to
observe two-photon processes.  However, several possibilities to
increase the harmonic fluxes are under investigation, based on
shortening the duration of the pump pulses
\cite{Salieres:1998:SST,Schafer:1997:U,Chang:1997:GCX} or improving
the phase-matching\cite{Rundquist:1998:U}. Furthermore, initial states
prepared as coherent superpositions of two or more bound states can
potentially increase the harmonic generation efficiency in the part of
the spectrum we are interested in \cite{Watson:1996:U,Sanpera:1983:U}.
Another challenge to be faced in a real experiment will be the making
of multilayer mirrors with different bandwidths to vary the number of
selected harmonics without changing their relative phases.  In spite
of these obstacles, we think that such an experiment is not out of
reach and could be useful in asserting the existence of subfemtosecond
harmonic pulses in the cases of locked phases.

{\bf Acknowledgments.} A.S would like to thank Richard Ta\"\i eb for
very illuminating ideas concerning section IV and acknowledges
financial support from DGICYT (Spain) contract number PB95-0778-C02-02
and European TMR program (FMRX-CT96-0080). E.C. thanks Pascal
Sali\`{e}res for a careful reading of the manuscript and P.
Lambropoulos for clarifying discussions.

\section*{Tables}
\begin{table}[htbp]
  \begin{center}
    \leavevmode

\begin{tabular}{llllllll}
\multicolumn{8}{c}{\bf Matrix elements: $s\rightarrow p \rightarrow d$} \\ \hline
-- & {\bf 19} & {\bf 21} & {\bf 23} & {\bf 25} & {\bf 27} & {\bf 29} & {\bf
31} \\ \hline
{\bf 19} &-84.1 &-66.1 &-58.2 &-49.1 &-41.8 &-35.8 &-30.9 \\ 
\hline
{\bf 21} & -90.5 & -74.4 & -61.9 & -52.0 & -44.2 & -37.9 & -32.7 \\ 
\hline
{\bf 23} & -105.4 & -86.2 & -71.4 & -59.9 & -50.8 & -42.7 & -37.5 \\ 
\hline
{\bf 25} & -168.1 & -136.4 & -112.4 & -94.0 & -79.5 & -68.0 & -58.6 \\ 
\hline
{\bf 27} & {\bf 132.1} & {\bf 105.1} & {\bf 85.3} & {\bf 70.5} & {\bf 
59.1} & {\bf 50.2} & {\bf 43.1} \\ \hline
{\bf 29} & -20.6 & -19.9 & -16.9 & -14.5 & -12.5 & -10.9 & -9.5 \\ 
\hline
{\bf 31} & -183.0 & -151.7 & -125.3 & -105.2 & -89.5 & -76.8 & -66.6
\\ 
\end{tabular}
    \caption{2-photon transition matrix-elements as a function of the
  order of each photon.}
    \label{tab:elements}
  \end{center}
\end{table}

\section*{Figures}

\begin{figure*}[hf]
\epsfxsize=8cm \epsfbox{H2s_RPpowerCentral.epsi}
\caption{Same as figure \ref{fig:H2s_Power_Law_RP_TOT} but restricted to 
the central peak. Open circles: average value. The power law fit gives
an argument of 1.3.}
\label{fig:H2s_Power_Law_RP_Central} 
\end{figure*} 

\newpage

\end{document}